# Hyperlinks embedded in Twitter as a proxy for total external inlinks to international university websites

**Enrique Orduña-Malea***

EC3 Research Group. Instituto de Diseño y Fabricación, Camino de Vera s/n, Polytechnic University of Valencia (UPV), 46022 Valencia, Spain. E-mail: enorma@upv.es, Tel: **+34** 963879480

**Daniel Torres-Salinas**

EC3 Research Group. Centro de Investigación Médica Aplicada, Avenida Pio XII, 55, Universidad de Navarra, 31008, Pamplona, Spain. E-mail: torressalinas@gmail.com

**Emilio Delgado López-Cózar**

EC3 Research Group. Facultad de Comunicación y Documentación, Colegio Máximo de Cartuja s/n, Universidad de Granada, 18071, Granada, Spain. E-mail: edelgado@ugr.es

**Abstract:** This article analyzes Twitter as a potential alternative source of external links for use in webometric analysis because of its capacity to embed hyperlinks in different tweets. Given the limitations on searching Twitter's public API, we decided to use the Topsy search engine as a source for compiling tweets. To this end, we took a global sample of 200 universities and compiled all the tweets with hyperlinks to any of these institutions. Further link data was obtained from alternative sources (MajesticSEO and OpenSiteExplorer) in order to compare the results. Thereafter, various statistical tests were performed to determine the correlation between the indicators and the ability to predict external links from the collected tweets. The results indicate a high volume of tweets, although they are skewed by the presence and performance of specific universities and countries. The data provided by Topsy correlated significantly with all link indicators, particularly with OpenSiteExplorer (r=0.769). Finally, prediction models do not provide optimum results because of high error rates, which fall slightly in nonlinear models applied to specific environments. We conclude that the use of Twitter (via Topsy) as a source of hyperlinks to universities produces promising results due to its high correlation with link indicators, though limited by policies and culture regarding use and presence in social networks.

**Introduction**

Webometrics is a discipline focused on the "study of web-based content with primarily quantitative research methods for social science goals using techniques that are not specific to one field of study" (Thelwall, 2009). Within this field, the analysis of hyperlinks (especially directed links from site A to site B) is one of the most widely-used techniques (Park & Thelwall, 2003), because the existence of a link from one resource (relating to a person, article, company, organization, etc.) to another involves a relationship, although the nature of this relationship is difficult to determine because the motives that have led to the establishment of a link are varied. This has generated an extensive debate in the literature (Smith, 2003; Wilkinson, Harries, Thelwall & Price, 2003; Bar-Ilan, 2005; Seeber et al., 2012; Kenekayoro, Buckley & Thelwall, 2013).

Link analysis becomes more reliable when applied to academic environments, such as universities, where external links received by the websites of these institutions bear some correlation with their scientific activity (Li, Thelwall, Musgrove & Wilkinson, 2003), although the target of these links will not necessarily lead to scientific articles or academic material but to general information pages or research-related activities (Thelwall, 2002a).

One of the main limitations of this type of research -apart from the exact nature of the relationship behind the hyperlink- is related to the coverage and potential bias of the sources used to retrieve these links. This has resulted in different research fronts with the aim of studying the search engines and their behavior over time, among other issues (Bar-Ilan, 2004; Vaughan & Thelwall, 2004; Thelwall, 2008).

Initially, the search engine most widely used for webometric purposes was Altavista[1]. This was subsequently acquired by Yahoo! in 2003, which then developed its own search engine (Yahoo! Search), thus tipping the balance of the market towards its service, which became the most widely used in the literature due to -among other things- the "linkdomain" search command, which enabled all links received by a certain website to be collected.

Specifically, the following query could be used to determine the number of incoming links to a website:

<linkdomain:mit.edu -site:mit.edu>

This query could even determine the number of links coming from a specific source (in this case, the links received by Massachusetts Institute of Technology from Harvard University):

< linkdomain:mit.edu site:harvard.edu>

The automation of these queries, by using Yahoo! API (Application Programming Interface), meant that, for a while, link analysis on academic web environments could be performed on a global scale (Thelwall & Smith, 2002; Aguillo, Granadino, Ortega & Priego, 2006).

However, the increasingly widespread use of these queries, both by researchers and by a community of professionals dedicated to SEO (Search Engine Optimization), began to generate excessive bandwidth consumption in the search engines (Thelwall & Stuart, 2006). This fact -along with other strategic reasons- led to this command being removed from other search engines with global coverage such as Bing (then MSN) and Google, remaining only on Yahoo! and Exalead.

In 2009, Microsoft and Yahoo finally announced an economic (and technological) agreement whereby Bing (the Microsoft search engine) became the exclusive search engine for both companies (The Washington Post, 2009). As Bing provided no support for the "link:" and "linkdomain:" commands (Seidman, 2007), the result of the agreement between the two companies was the complete disabling, in November 2011, of these search functions, both at Bing and at Yahoo! This consequently had a negative impact on webometric studies, since its most potent indicator could not now be obtained through a global search engine.

*Alternatives for link analysis*

The disabling of Yahoo! commands spurred the scientific community into action in the search for valid alternatives that would allow further link analysis. Their efforts can be clearly divided into two different strategies: the search for alternative indicators and the search for alternative sources (Ortega, Orduña-Malea & Aguillo, 2014).

**a) Alternative indicators**

The first strategy employed to solve the problem of a lack of link data from Yahoo! was the use of alternative indicators which do not require link commands but do provide information about relationships among websites. In this case, the proposed indicators are called "Title mentions" and "URL mentions".

The "Title mention" indicator (Cronin, Snyder, Rosenbaum, Martinson & Callahan, 1998) quantifies the number of times a character string (which may correspond to an author, the title of an article or journal, an institution, etc.) appears in the results of a search engine, by using a specific query, for example:

<"Yale university" -site:yale.edu>

Meanwhile, the "URL mention" indicator is similar to the above, with the difference being that the string matches the URL of the resource, so essentially it is conceptually closer to the hyperlink indicator. For example:

<"yale.edu" -site:yale.edu>

In order to analyze whether these indicators could be used as substitutes for external links, Thelwall (2011) compared the number of URL mentions with the external links, proving that the former are much lower in quantity. Subsequently, Thelwall and Sud (2011) expanded on this early work, performing a correlation analysis between external links and Title and URL mentions, proving the high correlation between the three indicators.

Later, Thelwall et al. (2012) concluded that the most appropriate indicator for building link diagrams would be URL mentions, but warned that a filtering data process would be needed, while Vaughan and Yang (2012) also concluded that this indicator can serve as a substitute for external links, although less so than "Sites linking in" from Alexa, which is based on counting the number of sites that link instead of the number of individual links.

In all these studies the limitations of these substitute indicators is remarked upon. Yet, although the correlations of these mention indicators with external links are high, there are two main problems:

a) these indicators do not express transitivity (understood as hypermedia transfer from one page to another) as the link does, so conceptually they are measuring another dimension of mentions; b) the fact that they are both made up of character strings favors the appearance of noise, mainly generated by synonymy and polysemy of language (a URL, in addition to the function of providing access to a resource, will identify it unambiguously).

Given these limitations, Ortega, Orduña-Malea and Aguillo (2014) analyze these two indicators in the context of the Spanish university system, identifying additional problems arising from language (Spain has an academic system with different official languages) and certain university policies on the Web (for example the use of multiple domains for one university). The conclusions of these authors point to the fact that both the Title and the URL mention positively correlate with external links, but that errors will lead to variance, according to each university system. They would not therefore be suitable for global analysis.

**b) Alternative link sources**

After the demise of Yahoo! Search (and the associated platform, Yahoo! Site Explorer), Exalead became the only search engine which supports link search commands, but its limited geographical coverage prevents it being used globally (Orduña-Malea, Serrano-Cobos, Ontalba-Ruipérez & Lloret-Romero, 2010).

On the other hand, new services are appearing which supply the links received by a particular website; these include MajesticSEO[2], Ahrefs[3] and Open Site Explorer.[4] Although the first two are employed in webometrics (currently used as sources of links in the Ranking Web of Universities)[5], the use of these platforms is very scarce in scholarly literature. The main problem with these products is that they are not general search engines (although Majestic SEO is evolving in this direction) and, moreover, they do not allow you to select the source of the links, or at least to automate this task.

Furthermore, the use of these platforms is hampered by the fact that they are paid services (allowing only a small number of queries per day or user), and access to the API of these services requires a service contract whose cost varies depending on the queries you wish to perform per day or month.

Another interesting service is Blekko[6], which was studied by Smith (2012) in a webometric analysis of institutional repositories. Although the results were not completely satisfactory, they opened the door to its future use, although at present, the Blekko SEO Tools[7] service no longer allows the registration of new users.

Recently, Vaughan (2012) proposed using Alexa[8] to obtain information about external links, in this case focusing interest on the "Sites linking in" tool, as mentioned previously. This indicator entails an aggregation of external links since all links from the same subdomain are only counted once. This concept had already been proposed by Thelwall (2002b) who, using ADM (Alternative Document Model), obtained good results and eliminated certain problems associated with the pure quantification of links (e.g. spam or a large number of links from a domain). Vaughan and Yang (2012) applied this indicator to the analysis of two largely significant samples (companies and universities), and concluded that it is a good replacement for the indicator that used to be provided by Yahoo!.

However, a number of problems have been identified when using this source, as Alexa obtains its data from a specific user base (those who have installed the Alexa toolbar in their browser). This skews representation per country of the sample, as it is more widely used in the English-speaking world.

On the other hand, MajesticSEO, Ahrefs and Open Site Explorer all offer information about domain-level links (called "referring domains" or "total linking root domains"), making them more suitable than Alexa for obtaining this indicator.

Despite attempts to explore and test new general sources and indicators that might serve as substitutes for global link metrics for webometric studies in academic environments, there has been insufficient analysis of social platforms (especially Twitter) for that purpose, inasmuch as they also permit the inclusion of hyperlinks. It is worth mentioning, at this point, a previous study carried out by Orduña-Malea and Ontalba-Ruipérez (2013), in which they identified the increasing weight of links from social platforms in the total number of external links that Spanish university websites received. They noted that the number of links from Wikipedia or Delicious to university websites could be used instead of the total external links received, given the high correlation found among them. And a still low but growing use may be observed on Twitter.

Notwithstanding, these analyses are scarce, and the use of these platforms for metric purposes is becoming more widespread in bibliometrics, where they are generating an intense academic debate (Priem & Hemminger, 2012; Priem, Piwowar, & Hemminger, 2012), because the analysis of the patterns of creation, dissemination and consumption of scholarly information in all these new web environments provides a more comprehensive and complete vision of the concept of impact as pertaining to academic work (Wouters & Costas, 2012; Torres-Salinas, Cabezas-Clavijo, & Jiménez-Contreras, 2013). This new approach was crystallized in the newly-coined Altmetrics (Priem, Taraborelli, Groth & Neylon, 2010).

*Scholarly use of Twitter*

Among the social platforms used for metric purposes, Twitter, the microblogging service created in 2006, stands out in particular surpassing 500 million users in 2012.[9]

The basic function of Twitter is to allow users to create messages of up to 140 characters (called tweets) and share them with other users. The expansion of Twitter, the tenth most visited website in the world in 2013 (according to Alexa[10]), has created a huge global deposit of tweets, the analysis of which is a challenge for researchers. Proof of this is the decision of the Library of Congress to create a future tweets archive to promote scholarly analysis of its contents.[11]

The messages or tweets may include complementary content that makes reference to:

- Internal material or material available within Twitter itself. A distinction may be made between, on the one hand, references to users and terms through the use of special commands ("@" and "#") that transform a text string into a hyperlink to other tweets that include the same labeled terms, and on the other hand, references to entire tweets through the use of retweets (Boyd, Golder, & Lotan, 2010).

- External material or material available outside of Twitter; in this case mainly pictures, videos and hyperlinks to other resources.

Although the motivations for using Twitter in professional and scholarly activities is a wide-ranging and complex issue (Zhao & Rosson, 2009), the analysis of informal communication between researchers through this tool is a topic of growing interest, as evidenced by the work of Priem, Costello, and Dzuba (2011), which examines the use of Twitter among scholars, or that of Letierce, Passant, Decker, and Breslin (2010), which studies the use of Twitter to communicate and disseminate scholarly messages.

Since URLs embedded in tweets may be considered acts of citation (Weller & Peters, 2012), the interest of bibliometrics in analyzing Twitter has not been long in coming. Thus, Priem and Costello (2010) define citations on Twitter as: "direct or indirect links from a tweet to a peer-reviewed scholarly article online." These authors analyzed the tweets of a group of researchers and observed the inclusion of links to peer-reviewed articles, while Weller, Dröge, and Puschmann (2011) subsequently extended this definition by considering that the hyperlink is only one way to reference scholarly papers in a tweet (mentions and retweets should be included), directly connecting traditional webometrics (link analysis) with altmetrics.

Weller and Puschmann (2011), analyzing a set of approximately 600 Twitter users (identified as researchers), observed high levels of citations in tweets related to scholarly activity. Later, Eysenbach (2011) completed a seminal work on the ability of Twitter to predict future citations. For this purpose, he collected 1,573 tweets with links to an article from the "Journal of Medical Internet Research (JMIR)" between 2009 and 2010, comparing these figures with citation data to said articles (17 to 29 months later) obtained from Scopus and Google Scholar. The results allowed the author to conclude that tweets can predict (with a low margin of error) highly cited articles from the number of tweets (with a link to the PDF of the article) generated during the three days following publication of the article.

Subsequently, Shuai, Pepe, and Bollen (2012) analyzed, among other altmetric indicators, the Twitter mentions of preprints deposited in ArXiv repository, and Haustein, Peters, Sugimoto, Thelwall & Lariviere (in press) discussed the use of Twitter in the dissemination of articles in the fields of Biomedicine and Health Sciences and concluded that there is a certain relationship between citations and tweets, although the latter also depend on the respective journals and areas of specialization. Thelwall, Haustein, Larivière and Sugimoto (2013) also noted the existence of a certain association between citations and tweets, although the real significance of this relationship requires further investigation, and may depend on the research discipline (Holmberg & Thelwall, 2013).

*Twitter as a new inlink source*

Although Twitter has now become an object of research in bibliometrics (as shown in the aforementioned studies), it is not so to the same extent purely from the webometrics perspective, despite the fact that link analysis is the core of this discipline. Links from social platforms to academic papers are being analyzed by altmetrics, but the study of links from these platforms to academic websites (especially universities) has scarcely been touched on by webometrics.

There have been some ranking products which list universities according to the performance of their corresponding Twitter user accounts, such as Track Social-universities[12] or Klout-universities.[13] The analysis of the Russell Group Universities in the UK is also worth noting, where each institutional Twitter account was measured by the number of followers, following, tweets, and tweets per month.[14] This report was later expanded on[15] with data gathered from Klout (Score, Network influence, amplification probability and True reach), Peerindex (Score, Activity, Audience, Authority) and Twitalyzer (Impact).

Nonetheless, these studies are focused on the analysis of Twitter accounts, taking into consideration indicators related to them, but there is a lack of studies centered on the use of Twitter as an alternative source of link data for webometric purposes.

One of the possible reasons for the lack of this kind of study is the difficulty in obtaining thorough metrics because the Twitter public API has coverage limitations when collecting tweets, i.e. it does not recover all existing tweets for a specific query (Reips & Garaizar, 2011; Bogers & Björneborn, 2013). Although various interesting tools do exist (such as BackTweets Peerindex TweetStats, Twitalyzer and Social mention, among others), the most comprehensive data source is probably Topsy[16], as it currently has more than 425 billion tweets indexed (every public Twitter message since the first tweet was posted in 2006)[17], and its tweet archive is even greater than the Twitter public API.

Nonetheless, the use of Twitter (and Topsy) for webometric purposes to quantify the number of links from tweets to university websites in order to check their correlation with total external links (a key indicator in the Ranking Web of Universities) is a topic not yet covered in webometrics literature.

*Objectives*

The main objective of this paper is to analyze the possible use of Twitter as an alternative source of hyperlinks for webometric studies applied to universities.

To this end the following specific objectives were considered:

1) To quantify the number of tweets with embedded links to a sample of 200 international university websites.

2) To compare this indicator with various external link sources, checking if university performance clusters obtained are similar, independently of the source.

3) To determine whether it is possible to predict the total amount of external links from the number of tweets.

**Methodology**

*Information sources and indicators*

First, the sources were identified, as was the method of obtaining both tweets and other web indicators (external links).

To this end, the following sources were consulted: Topsy (number of accumulated tweets: full, month, week), Majestic SEO (external links, referring domains) and Open Site Explorer (external links, referring domains). Ahrefs was not used because of its inferior coverage.

Topsy, created in 2007 by Topsy Labs[18], is designed to collect and index tweets and then position them in terms of their impact (measured, amongst other factors, by the number of links received, retweets or times that they have been marked as favorites).

Topsy also provides complementary value-added services, such as searching for specific material (videos, images, links), comparative graphs based on various metrics, as well as coverage of other platforms (specifically Google+).

For specific queries, the "site:domain" command was used; this returns the number of tweets in which an embedded hyperlink to the corresponding domain appears (not to be confused with the "site:" command used by general search engines like Google). This system retrieves both the cumulative total of tweets and those accumulated during the last month and the last week. For this study, the three metrics were obtained for each of the 200 universities.

Topsy was used instead of the Twitter public API due to its higher coverage of tweets. As previously discussed, the Twitter API does not support searching of all public tweets. Therefore, for the purposes of this research (to quantify the number of tweets with links to universities, and not the actual Tweet contents), Topsy was more appropriate. Although the Twitter public API can provide additional information for the corresponding tweets (such as number of retweets, among others), this is not important for the goals of this research.

Figure 1 shows the Topsy search page, the "site:" command and the gathering of metrics and results. It shows how the system identifies URLs embedded in every tweet collected (either full or short URLs), which may be web pages (HTML, ASP, PHP, etc.) or office files (DOC, PPT, PDF, etc.), and whose position in the results page is due to criteria related to the impact of the tweet in social networks, which is not pertinent to this study.

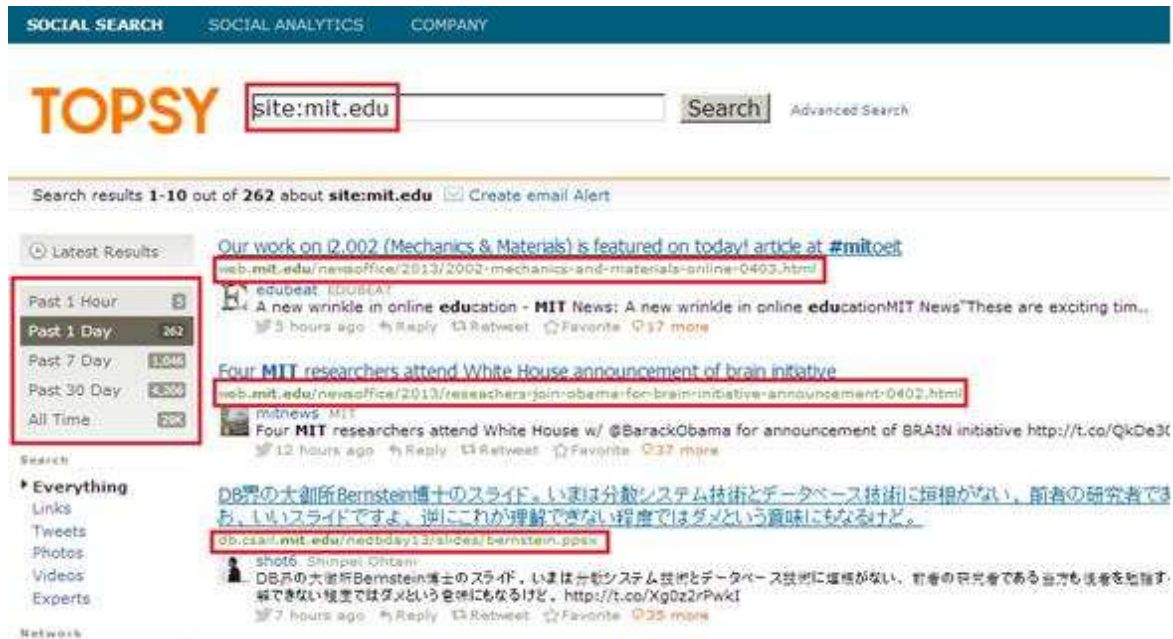

**FIG 1. Query in Topsy of tweets with embedded hyperlinks that point to a specific academic website.**

Majestic SEO is a comprehensive database of links created in 2008 by Majestic-12 Ltd, whose historic index currently covers more than 2 trillion URLs. The company also offers value-added services (both free and paid), one of which, the generation of Flow Metrics, stands out in particular. The system provides access to a wide range of web performance indicators through a simple URL search, as well as direct API requests.

Open Site Explorer is an online application created in 2010 to extract and analyze link data from the Mozscape Index, a data repository (and associated toolset) that provides access to over 80 billion URLs, created in 2008 by SEOMoz (now MOZ). It also provides free and paid services, search engine or API requests, and generates its own metrics (MozRank, MozTrust and Domain Authority).

Table 1 shows in summary the different indicators used in this study and a brief description of each.

**TABLE 1. Definitions of indicators and sources.**

| INDICATOR | SOURCE | DEFINITION |
|---|---|---|
| Full Tweets | Topsy | Number of tweets (contained in the Topsy database) with at least 1 embedded URL which points to a university web domain |
| Month Tweets | Topsy | Number of tweets (contained in the Topsy database) with at least 1 embedded URL which points to a university web domain, created last month |
| Week Tweets | Topsy | Number of tweets (contained in the Topsy database) with at least 1 embedded URL which points to a university web domain, and created last week |
| MAJ (Links) | Majestic | External inlinks (backlinks): total number of external web pages containing links to a university web domain |
| MAJ (Sites) | Majestic | Referring domains: number of domains from which a backlink is pointing to a university web domain |
| OSE (Links) | Open Site Explorer | External inlinks: number of links that come from external domains to a university web domain |
| OSE (Sites) | Open Site Explorer | Linking root domain (referring domains): number of unique root domains containing at least one link to a university web domain |

It should be noted that the OSE (Links) and MAJ (Links) indicators are not completely equivalent. Although both are based on quantifying the number of external links received by a web domain, MAS (Links) focuses on quantifying the number of pages that contain a link to another page. In other words, if page A contains 3 links to page B, only 1 backlink is counted whereas with OSE (Links), if page A contains 3 links to page B, then 3 backlinks are counted. OSE (Sites) and MAJ (Sites) indicators, however, are equivalent.

*The study sample*

Second, the sample of universities, to which the various, previously established indicators were to be applied, was selected. Since the objective was to analyze a representative set of universities with high performance and web visibility (hyperlinks received from Twitter), the top 200 universities according to the Ranking Web of Universities (January 2013 edition) were selected.[5] This ranking has been widely used as a sample of university websites for webometric analysis as it covers a comprehensive collection of institutions worldwide (Aguillo, Granadino, Ortega & Prieto, 2006), which are measured by page count and external inlink indicators.[5]

For each of these universities the URL of the official website and the corresponding country were compiled.

*Statistical analysis of the sample*

In all three sources (Topsy, MajesticSEO and Open Site Explorer), the search and collection of indicators were performed manually at the exact same time (without searching the API) during the month of March 2013. The results were transferred to a spreadsheet for statistical analysis, using the XLSTAT software suite.

In order to provide answers to our previously defined specific objectives 2 and 3, we conducted the following statistical analysis of the data:

First, we calculated the correlations between the seven indicators analyzed (See Table 1). Since link distribution follows a power distribution pattern, the Spearman correlation coefficient was chosen to perform calculations. Later, statistical tests were carried out to ascertain the similarities and differences between the sources and indicators. The tests performed were:

- Principal Component Analysis (PCA): applied in order to complement correlation analysis by finding causes that explain the variability of the indicators applied to the sample. The Pearson(n) PCA with varimax rotation was applied.

- Cluster analysis (k-means): aims to partition "n" observations into "k" clusters in which each observation belongs to the cluster with the nearest mean, serving as a prototype of the cluster.

- Friedman test for "n" samples in conjunction with the Nemenyi post-hoc test: studies the statistical differences between the web indicators and the different sources that provide these data. This is a non-parametric test, which detects differences between paired samples. It was used in combination with the post-hoc test, which points out the samples which differ between themselves.

- Kruskal-Wallis test: detects whether or not "n" datasets belong to the same population. This statistical method is a suitable non-parametric test for non-normal distributions such as web data.

Second, predictive models were designed for the purpose of checking whether the values achieved by Topsy could be used to predict the external links in the other sources. To this end, two types of regression models were performed.

One was a linear regression (after transforming the raw data into logarithmic values) and the other was a nonlinear regression model, for the purpose of determining the degree of precision, as compared to the linear regression model, with respect to link distribution.

The complete raw data for the 200 universities analyzed, both from Topsy and the other sources used (Majestic SEO and Open Site Explorer) are available in the supplementary material, disaggregated by university (for each corresponding website) and indicator.[19]

**Results**

First, there is a descriptive analysis of the presence in Topsy of the 200 universities sampled. Next, both the correlation between the seven web indicators and the similarity between universities based on performance in Topsy (through the identification of clusters) are analyzed. Finally, linear and nonlinear regression models are applied in order to estimate the ability of Topsy to predict the total number of links received by universities.

*Descriptive analysis of the presence of universities in Topsy*

In order to describe the total volume of data that Topsy is able to retrieve in the sample, the number of tweets retrieved (weekly, monthly and total) is shown below, as well as other significant statistical values (Table 2).

**TABLE 2. General statistics for tweets gathered by Topsy for the university sample**

| INDICATOR | TOTAL | MEAN | DEVIATION | MEDIAN | MAX | MIN |
|---|---|---|---|---|---|---|
| Full | 1,212,741 | 6,063 | ± 20,742 | 3,040 | 286,775 | 3 |
| Month | 177,534 | 887 | ± 1,971 | 508 | 25,419 | 0 |
| Week | 45,400 | 227 | ± 517 | 121 | 6,768 | 0 |

The total number of results retrieved with a hyperlink to any of the 200 universities in the sample is 1,212,741 tweets, although the data show a very uneven distribution, which will be covered in greater detail in later sections.

The high percentage of tweets per week (3.74%) and month (14.6%) in relation to the total reflects both an exponential growth in tweets and a historical coverage that was not complete at the time of the search, a factor that affects all universities equally, thus minimizing its effect.

Table 3 lists the 20 universities with the highest total number of tweets recorded by Topsy. Additionally, results for monthly and weekly records are given, together with the percentage of the total that they represent. Finally the position of each university in the Webometric Ranking (WR) is included.

**TABLE 3. University ranking according to the number of tweets gathered by Topsy.**

| RANK | UNIVERSITY | COUNTRY | Full | Month | Week | Month (%) | Week (%) | WR |
|---|---|---|---|---|---|---|---|---|
| 1 | * Universidad Nacional Autónoma de México | MX | **286,775** | 25,419 | 6,768 | 8.86 | 2.36 | 36 |
| 2 | Iowa State University | US | **44,119** | 7,863 | 1,562 | 17.82 | 3.54 | 59 |
| 3 | Harvard University | US | **34,039** | 4,764 | 1,362 | 14.00 | 4.00 | 1 |
| 4 | Massachusetts Institute of Technology | US | **26,818** | 4,052 | 1,126 | 15.11 | 4.20 | 3 |
| 5 | Stanford University | US | **25,798** | 3,885 | 1,133 | 15.06 | 4.39 | 2 |
| 6 | Princeton University | US | **22,844** | 4,355 | 999 | 19.06 | 4.37 | 24 |
| 7 | University of Texas Austin | US | **19,721** | 2,076 | 418 | 10.53 | 2.12 | 12 |
| 8 | ** Pennsylvania State University | US | **18,554** | 2,712 | 712 | 14.62 | 3.84 | 11 |
| 9 | University of Pennsylvania | US | **17,237** | 2,371 | 712 | 13.76 | 4.13 | 5 |
| 10 | University of Wisconsin Madison | US | **13,767** | 2,162 | 598 | 15.70 | 4.34 | 18 |
| 11 | University of Minnesota | US | **13,213** | 2,038 | 569 | 15.42 | 4.31 | 10 |
| 12 | University of California Berkeley | US | **12,655** | 2,145 | 614 | 16.95 | 4.85 | 7 |
| 13 | University of British Columbia | CA | **12,576** | 1,875 | 519 | 14.91 | 4.13 | 22 |
| 14 | University of Michigan | US | **12,208** | 1,948 | 511 | 15.96 | 4.19 | 4 |
| 15 | Yale University | US | **12,024** | 1,597 | 428 | 13.28 | 3.56 | 13 |
| 16 | University of Oxford | UK | **11,988** | 2,194 | 500 | 18.30 | 4.17 | 16 |
| 17 | Columbia University New York | US | **11,841** | 2,022 | 498 | 17.08 | 4.21 | 8 |
| 18 | University of Tokyo | JP | **11,350** | 1,723 | 436 | 15.18 | 3.84 | 48 |
| 19 | London School of Economics and Political Science | UK | **11,140** | 2,208 | 652 | 19.82 | 5.85 | 163 |
| 20 | Cornell University | US | **10,512** | 1,853 | 471 | 17.63 | 4.48 | 9 |

*Webpages from the newspaper "La Jornada" have been excluded from the UNAM figures.

**Data from CiteSeerX has been excluded from the calculations related to Pennsylvania State University.

The results are generally high: 22 universities (those included in Table 1, plus Michigan State University and the University of Cambridge) exceed 10,000 overall results (defined as the number of tweets with an embedded link to any URL belonging the university web domain).

The first position is unexpectedly occupied by the National Autonomous University of Mexico (UNAM), with total (286,775) monthly (25,419) and weekly (6,768) values well above the other universities. Second position is occupied, equally unexpectedly, by Iowa State University (44,119 results), and third position (Harvard University) obtained 34,039 results.

Although the leading universities in the WR have a high ranking in Topsy (Harvard, MIT, Stanford, among others), some institutions with a low ranking in the WR achieve higher positions in Topsy, like Iowa State University, the University of Tokyo, and especially the London School of Economics and Political Science (LSE) and UNAM. By contrast, other universities occupy lower positions, notably the University of California-Los Angeles (from sixth in the WR to twenty-fifth in Topsy). In any case, the overall correlation between the positions of the universities in the WR and their positions relative to the total number of tweets retrieved by Topsy is moderate ($r_p$=0.63; α=0.01).

With regard to the monthly and weekly percentage data, high levels are observed in these top positions. Specifically, of the 200 universities in the sample, monthly data do not equal at least 10% of the total for only 13 of them. Among these, there are six Chinese and four Brazilian universities, as well as UNAM, due to its high total recorded value. Other particularly high monthly values were recorded for the University of Chicago (31.28%), University of Edinburgh (23.31%), University of Auckland (21.67%), University of Miami (23.83%) and Università degli Studi di Pisa (28.43%), all of which totaled over 1,000 records.

With respect to distribution by country, the geographical distribution of the initial sample of 200 universities from the WR should be taken into account; this is indicated in Table 4 (column N), which shows that the United States (88) and Canada (14) together account for more than 50 percent, followed by Germany (13) and the UK (11).

**TABLE 4. University distribution by country according to the number of tweets in Topsy.**

| COUNTRY | N | TOP 100 | TOP 100 WR | TOP 50 | TOP 50 WR | COUNTRY | N | TOP 100 | TOP 100 WR | TOP 50 | TOP 50 WR |
|---|---|---|---|---|---|---|---|---|---|---|---|
| **United States** | 88 | 67 | 63 | 37 | **39** | Norway | 3 | 1 | 1 | 0 | **0** |

| Canada | 14 | 9 | 5 | 2 | 2 | Portugal | 3 | 0 | 1 | 0 | 0 |
|---|---|---|---|---|---|---|---|---|---|---|---|
| Germany | 13 | 0 | 1 | 0 | 0 | Taiwan | 3 | 0 | 1 | 0 | 1 |
| United Kingdom | 11 | 11 | 5 | 6 | 3 | Austria | 2 | 0 | 1 | 0 | 0 |
| China | 9 | 0 | 6 | 0 | 0 | Japan | 2 | 2 | 2 | 2 | 1 |
| Netherlands | 7 | 3 | 4 | 0 | 1 | Czech Republic | 1 | 0 | 1 | 0 | 0 |
| Australia | 6 | 0 | 1 | 0 | 0 | Greece | 1 | 0 | 0 | 0 | 0 |
| Sweden | 6 | 0 | 1 | 0 | 0 | Jerusalem | 1 | 0 | 0 | 0 | 0 |
| Spain | 5 | 5 | 1 | 1 | 0 | Mexico | 1 | 1 | 1 | 1 | 1 |
| Italy | 5 | 0 | 0 | 0 | 0 | New Zealand | 1 | 0 | 0 | 0 | 0 |
| Brazil | 4 | 1 | 1 | 1 | 1 | Russia | 1 | 0 | 1 | 0 | 0 |
| Switzerland | 4 | 0 | 1 | 0 | 1 | Singapore | 1 | 0 | 1 | 0 | 0 |
| Belgium | 3 | 0 | 1 | 0 | 0 | Slovenia | 1 | 0 | 0 | 0 | 0 |
| Denmark | 3 | 0 | 0 | 0 | 0 | Thailand | 1 | 0 | 0 | 0 | 0 |

Although the universities are positioned according to the number of total results in Topsy, there are some important differences. While the United States leads the ranking (37 of the 50 highest-scoring universities are North American), the values for Germany and China (and even Australia and Sweden) are particularly negative as none figures in the top 100 positions, despite their having many universities in the overall sample.

In addition, the number of countries in the top 50 and 100 of the WR is included, with the purpose of showing which countries are most positively or adversely affected by the fact that they are measured by number of tweets. Spain is particularly favored in Topsy (from 1 university in the top 100 in WR to 5 in the top 100 in Topsy), as are the UK (from 5 to 11 universities) and Canada (from 5 to 9).

Conversely, the trend is negative in the case of China (from 6 universities in WR to none in the top 100 in Topsy, as discussed previously). In the case of the United States, there is no significant change in the top 50, although in the case of the top 100 it slightly increases its presence in Topsy (from 63 to 67 universities).

*Similarity between indicators*

Table 5 shows the correlations obtained between the different web indicators analyzed. The results indicate the high correlation between total Topsy values and corresponding monthly (r=0.98) and weekly (r=0.97) values, due in part to the high percentage of the latter in relation to the former (see Table 3).

With respect to the correlations of Topsy (full) with the other indicators, they are somewhat lower, especially with MajesticSEO, while the maximum value obtained is between Topsy and OSE Sites (r=0.77). Unexpectedly, both MAJ and OSE both correlated higher with Topsy for referring domains than for external inlinks.

The correlations of Topsy (month) and Topsy (week) with the other indicators give figures very similar to those obtained by Topsy (full). This may indicate that the differences between universities in Topsy are detectable even in short time intervals (even weekly).

High correlations were also observed between external inlink and referring domain indicators obtained by the same source, both in MajesticSEO (r=0.82) and Open Site Explorer (r=0.81). The correlations between different sources are somewhat lower, but equally significant; particularly striking is the correlation between MAJ Sites and OSE Sites (r=0.93).

**TABLE 5. Correlation matrix between indicators and sources.**

| FUENTE | Topsy (full) | Topsy (month) | Topsy (week) | MAJ (Links) | MAJ (Sites) | OSE (Links) | OSE (Sites) |
|---|---|---|---|---|---|---|---|
| **Topsy (full)** | 1.00 | | | | | | |
| **Topsy (month)** | 0.98** | 1.00 | | | | | |
| **Topsy (week)** | 0.97** | 0.99** | 1.00 | | | | |
| **MAJ (Links)** | 0.58** | 0.58** | 0.54** | 1.00 | | | |
| **MAJ (Sites)** | 0.68** | 0.68** | 0.65** | 0.82** | 1.00 | | |
| **OSE (Links)** | 0.66** | 0.67** | 0.63** | 0.76** | 0.76** | 1.00 | |
| **OSE (Sites)** | 0.77** | 0.76** | 0.73** | 0.77** | 0.93** | 0.81** | 1.00 |

\*\* significant values (except diagonal) at the level of significance alpha=0.001 (two-tailed test)

Considering the uneven geographical distribution shown in Table 4, the greater presence of the United States and Canada could influence the values obtained. In order to test this effect, the correlations were recalculated considering only the universities from these two countries. The results indicate similar correlations, although a slight increase was detected, precisely in the values obtained for Topsy and MAJ Sites (r=0.77) and OSE Sites (r=0.84).

To shed light on the relationships between the web indicators used, a Principal Component Analysis (PCA) was performed, the results of which are shown in Table 6. The biplot (after varimax rotation) is displayed on the left, the aim of which is to clearly visualize the grouping of the various indicators, while the identified components are displayed on the right to show numerically the extent to which each of them contributes to the variance.

**TABLE 6. PCA Analysis: biplot and factor loadings after varimax rotation.**

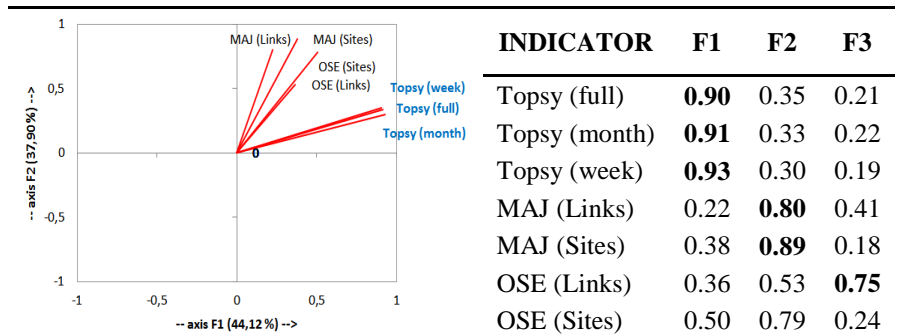

| INDICATOR | F1 | F2 | F3 |
|---|---|---|---|
| Topsy (full) | **0.90** | 0.35 | 0.21 |
| Topsy (month) | **0.91** | 0.33 | 0.22 |
| Topsy (week) | **0.93** | 0.30 | 0.19 |
| MAJ (Links) | 0.22 | **0.80** | 0.41 |
| MAJ (Sites) | 0.38 | **0.89** | 0.18 |
| OSE (Links) | 0.36 | 0.53 | **0.75** |
| OSE (Sites) | 0.50 | 0.79 | 0.24 |

Bold figures: major factor loadings found.

Seven components were identified (which corresponds to the number of indicators), which were reduced to 3 after the varimax rotation, the first two of which (F1 and F2) explain 82.01% of the variance (Figure 2). The data indicate high percentages for the 3 Topsy indicators in F1, for Majestic SEO (links and sites) in F2, and finally, for OSE (links) in F3.

The values in Table 6, apart from the explained variances, confirm the previous correlations. On the one hand, it shows the components associated with Topsy and, on the other hand, the components related to links and sites, which are closer to the Topsy data, especially OSE, which reinforces the idea that Topsy is a universe with its own characteristics when obtaining hyperlinks, slightly different from the other sources.

To statistically determine these differences between sources and indicators, the Kruskal-Wallis test (H=1316.784; *p*-value=<.0001) and the Friedman test (Q=1197.984; *p*-value=<.0001) were performed. The conclusion in both is a rejection of the null hypothesis of no difference between the seven samples (i.e. the difference between samples does have significance).

In addition, the Nemenyi post-hoc test was applied. This also identified seven groups (one per indicator), confirming the statistically significant difference between the seven indicators obtained. In this case, even the indicators from the same sources (Topsy, MAJ and OSE) appear in separate groups. This further reinforces the importance of the representative correlations in Table 5.

*Similarity between universities*

The information shown in Table 6 is further developed in Figure 2, in which, in addition to identifying the components, adds the qualitative variable relating to the geographic location of each of the universities.

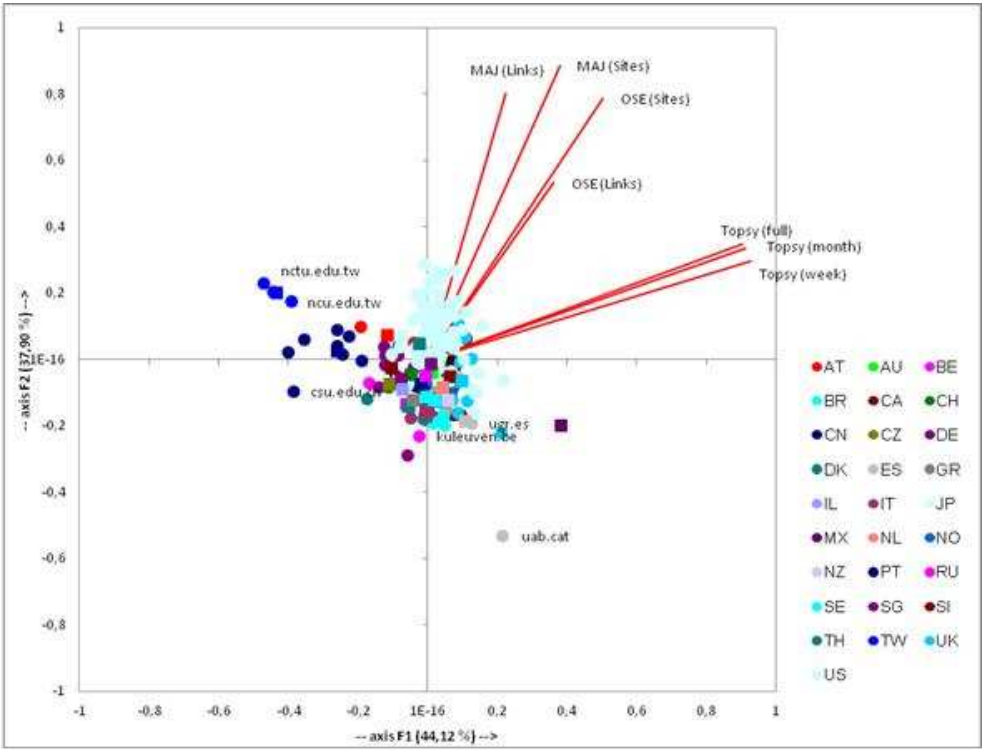

**FIG 2. Principal Component Analysis (PCA) of the sample of universities (Topsy, MAJ and OSE).**

The data show how US universities are grouped almost entirely in the first quadrant (top right), while the Chinese and Taiwanese universities appear in the fourth quadrant (top left). This indicates that the latter universities perform well in link indicators but are clearly underrepresented in Topsy (Annex I), since, of the 15 universities with the fewest total tweets, there are 8 Chinese (of a total of 9) and all the Taiwanese universities (3), aspects that are partly commented on in Table 4.

Given the differences between universities and their distribution in the different quadrants of the biplot for the first two components, a k-means clustering process was performed (k= 5; Annex II), taking into account the values obtained for the 7 indicators considered. The first two clusters obtained are smaller in size (10 and 14 universities respectively), and the remaining clusters are larger (43, 63 and 70 universities).

The first cluster is made up exclusively of 10 US universities, including the best positioned in the WR, albeit with some significant absences (University of Pennsylvania, University of California-Los Angeles and Columbia University, amongst others) and some institutions from outside the WR top ten, albeit in high positions, such as the University of Washington (23rd) and Michigan State University (26th). The second cluster, on the other hand, is composed almost entirely of Chinese (8) and Taiwanese (3) universities. The third cluster contains virtually all the US universities not included in the first cluster, together with non-US universities that perform extremely well on Twitter (mainly Oxford, Cambridge and UNAM). Finally, the fourth cluster is characterized by the fact that it includes most of the German (12) and Swedish (6), and the fifth, most of the British (8) universities.

*Models predicting the number of links*

After the analysis of the correlation between university indicators and clusters as described in the preceding paragraphs, various regression models were then applied in order to determine the ability to predict links received from the number of tweets.

First, a linear regression model was carried out, taking the OSE (Sites) indicator as an explanatory variable, as it obtained the highest correlation values with Topsy (full), the variable to be explained. The results are shown in Figure 3, which shows that the correlation is high, although there are several universities acting as outliers, which clearly influences the low coefficient of determination value ($r^2$= 0.39).

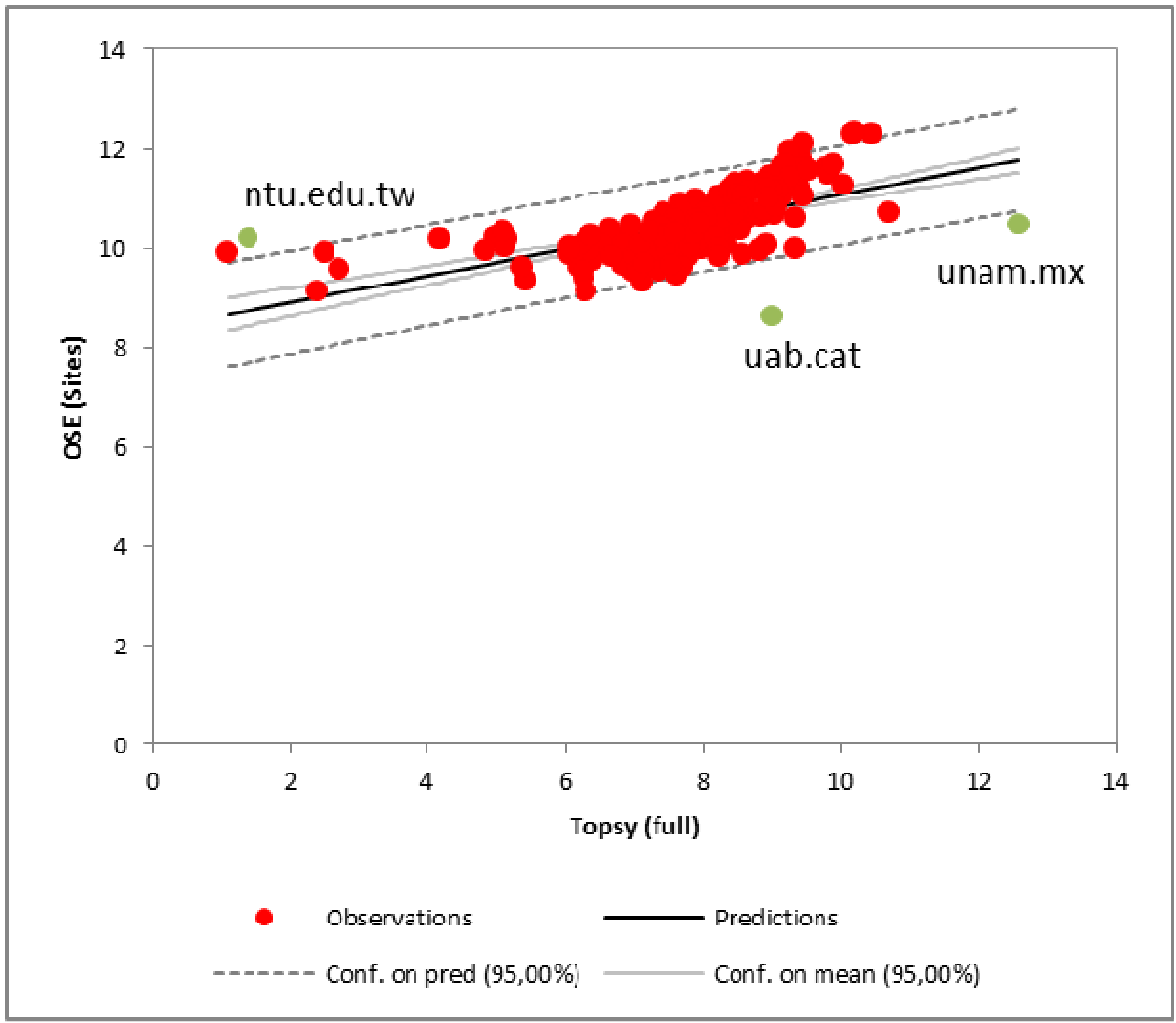

**FIG 3. Linear regression model (top 200).**

The equation of the model obtained with the non-standard coefficients is shown below:

   OSE (Sites) = 8.37 + 0.27*Topsy (full);

Since the distribution of links follows a power law, linear regression may not be the most appropriate, even though the values were logarithmically transformed. In this case, to check possible differences and improvements in the predictive model, an additional nonlinear regression was performed. Table 7 shows the nonlinear functions that yield a higher coefficient of determination, together with the values of the constant parameters.

**TABLE 7. Coefficient of determination for different nonlinear models.**

| Model | R² | SSR |
|---|---|---|
| pr1+pr2*X1^1+pr3*X1^2 | 0,479 | 42,704 |
| pr1*Exp(pr2*X1) | 0,403 | 48,889 |
| pr1*Exp(pr2*X1)+pr3 | 0,444 | 45,508 |
| pr3/(1+Exp(-pr1-pr2*X1)) | 0,402 | 48,962 |
| **pr1+(pr4-pr1)/(1+(X1/pr3)^pr2)** | **0,586** | **33,881** |
| Exp(pr1+x1*pr2)/(X1 + 1)^pr3+Exp(pr4+X1*pr5)/(X1 + 1)^pr6 | 0,487 | 41,988 |
| pr1+pr2*Cos(2*Pi*pr3*X1)+pr4*Sin(2*Pi*pr3*X1)) | 0,479 | 42,674 |

| Parameters | Value | Standard deviation |
|---|---|---|
| pr1 | 11,661 | 0,214 |
| pr2 | 13,982 | 2,998 |
| pr3 | 8,546 | 0,179 |
| pr4 | 9,873 | 0,076 |

Then the model with the best coefficient ($r^2$ = 0.59) was chosen, to display the nonlinear regression (Figure 4).

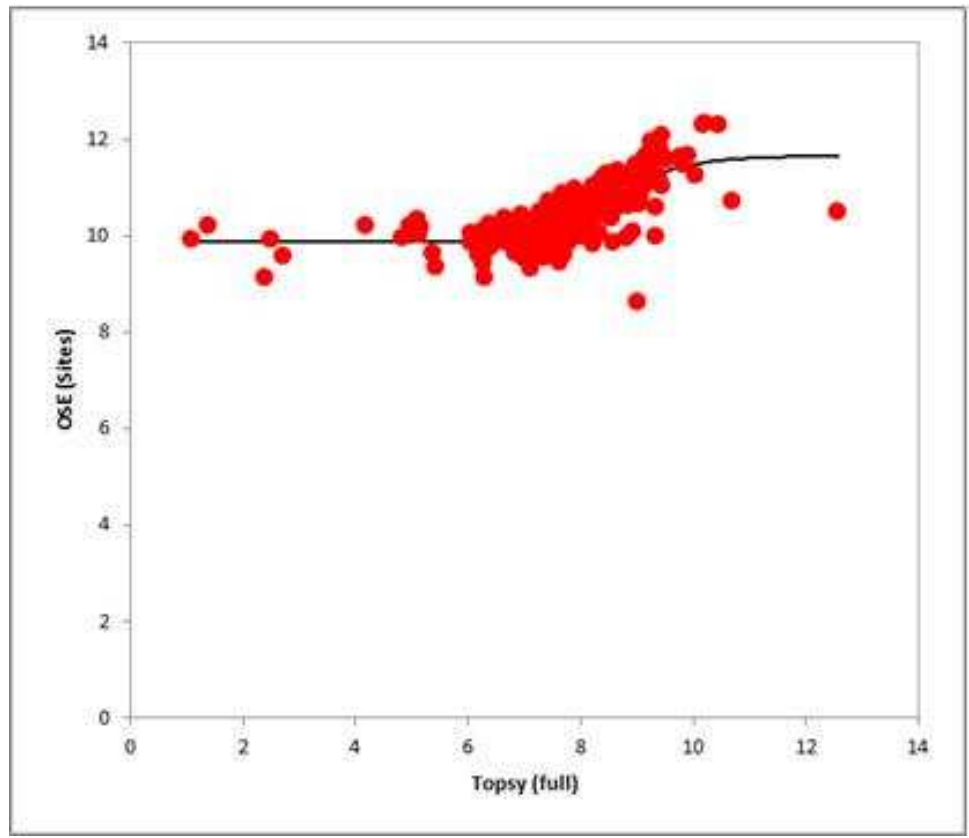

**FIG 4. Nonlinear regression model (top 200).**

The results show a more accurate regression model than the linear model, with a higher coefficient of determination, despite the fact that universities with extreme results continue to feature.

Finally, after obtaining the two predictive models (linear and nonlinear), they were applied to the sample to check its accuracy with the real data that had been obtained.

The results are shown in Table 8, which provides, for the first 20 universities according to Topsy (full), OSE (sites) values obtained from searching this source directly, and those obtained from predictive models (model), together with the corresponding error rate. The data for all the universities is in Annex 3 of the supplementary material.

**TABLE 8. Application of predictive models (top 20).**

| WEB DOMAINS | Topsy (full) | NONLINEAR REGRESSION | | | | LINEAR REGRESSION | | | |
|---|---|---|---|---|---|---|---|---|---|
| | | OSE | OSE (model) | Dif | Dif (%) | OSE | OSE (model) | Dif | Dif (%) |
| unam.mx | **286,775** | 35,822 | 114,966 | 79,144 | 220.9 | 35,822 | 127,816 | 91,994 | 256.8 |
| iastate.edu | **44,119** | 44,584 | 107,584 | 63,000 | 141.3 | 44,584 | 77,142 | 32,558 | 73.0 |
| harvard.edu | **34,039** | 218,695 | 104,536 | -114,159 | -52.2 | 218,695 | 71,929 | -146,766 | -67.1 |
| mit.edu | **26,818** | 228,550 | 100,808 | -127,742 | -55.9 | 228,550 | 67,448 | -161,102 | -70.5 |
| stanford.edu | **25,798** | 216,825 | 100,104 | -116,721 | -53.8 | 216,825 | 66,746 | -150,079 | -69.2 |
| princeton.edu | **22,844** | 78,215 | 97,700 | 19,485 | 24.9 | 78,215 | 64,592 | -13,623 | -17.4 |
| utexas.edu | **19,721** | 118,076 | 94,373 | -23,703 | -20.1 | 118,076 | 62,081 | -55,995 | -47.4 |
| psu.edu | **18,554** | 96,791 | 92,851 | -3,940 | -4.1 | 96,791 | 61,067 | -35,724 | -36.9 |
| upenn.edu | **17,237** | 111,783 | 90,899 | -20,884 | -18.7 | 111,783 | 59,867 | -51,916 | -46.4 |
| wisc.edu | **13,767** | 104,720 | 84,172 | -20,548 | -19.6 | 104,720 | 56,344 | -48,376 | -46.2 |
| umn.edu | **13,213** | 108,387 | 82,822 | -25,565 | -23.6 | 108,387 | 55,723 | -52,664 | -48.6 |
| berkeley.edu | **12,655** | 178,121 | 81,366 | -96,755 | -54.3 | 178,121 | 55,079 | -123,042 | -69.1 |
| ubc.ca | **12,576** | 61,832 | 81,152 | 19,320 | 31.2 | 61,832 | 54,986 | -6,846 | -11.1 |
| umich.edu | **12,208** | 132,435 | 80,124 | -52,311 | -39.5 | 132,435 | 54,547 | -77,888 | -58.8 |
| yale.edu | **12,024** | 110,765 | 79,593 | -31,172 | -28.1 | 110,765 | 54,324 | -56,441 | -51.0 |
| ox.ac.uk | **11,988** | 75,421 | 79,487 | 4,066 | 5.4 | 75,421 | 54,280 | -21,141 | -28.0 |
| columbia.edu | **11,841** | 109,529 | 79,051 | -30,478 | -27.8 | 109,529 | 54,100 | -55,429 | -50.6 |
| u-tokyo.ac.jp | **11,350** | 39,838 | 77,534 | 37,696 | 94.6 | 39,838 | 53,485 | 13,647 | 34.3 |
| lse.ac.uk | **11,140** | 21,941 | 76,856 | 54,915 | 250.3 | 21,941 | 53,216 | 31,275 | 142.5 |
| cornell.edu | **10,512** | 156,790 | 74,713 | -82,077 | -52.3 | 156,790 | 52,390 | -104,400 | -66.6 |

From the results obtained after applying the two models, very high error rates in the prediction of linked sites may be observed. Assuming a tolerable error rate of 10%, due to the variability of search engine data and rounding processes, only 46 universities (23%) in the nonlinear model and 42 (21%) in the linear model have error rates lower than 10%.

In some cases, the high performance of universities in Twitter (as is the case of UNAM, Iowa State University, LSE or the Autonomous University of Barcelona) causes the predictive models to estimate high numbers of websites, much lower than those actually obtained, while in other cases the reverse happens: a lower performance on Twitter leads to underrepresentation in the predictive model, which is precisely what happened to the group of Chinese and Taiwanese universities.

These two groups of universities (underrepresented and overrepresented in Twitter) may affect the accuracy of the model, and this could be having an indirect influence on the values obtained for the top ranked universities in the WR, such as MIT, Harvard and Stanford which, despite their high performance on Twitter, tend to be underrepresented by the models used.

For this reason, we decided to replicate the models, applying them only to the US and Canada, to see if their accuracy is higher. In this case, the 102 respective universities were considered, and the following models were obtained:

Linear:

OSE (Sites) = 6.01 + 0.57*Topsy (full);

$r^2$: 0.65

Nonlinear:

OSE (Sites) = pr1+pr2*Cos(2π*pr3*X1)+pr4*Sin(2π*pr3*X1));

Parameters: pr1: 11.059; pr2: 0.648; pr3: -0.193; pr4: 0.256;

$r^2$: 0.71

The coefficients of determination obtained are appreciably higher than those achieved with the overall sample of 200 universities. Observed error rates also improved compared to the data presented in Table 8, especially for the new linear model. For example, the error of underrepresentation in the estimated number of websites for Harvard fell from 67.1% to 26%; MIT, from 70.5% to 38.2%; and Stanford, from 69.2% to 36.3%. Complete data for these 102 universities are available in the supplementary material (Appendix IV).

Furthermore, since the sample was more specific, the errors caused by extreme values were exacerbated. For example, Iowa State University (the error rate does improve in the linear model, but worsens in the nonlinear model) and Princeton University (worse in both models). Even so, the overall results are clearly superior, giving 33 universities (32.3%) with less than 10% error in the nonlinear model and 30 in the linear model.

Figure 5 shows the regression curves obtained for both models.

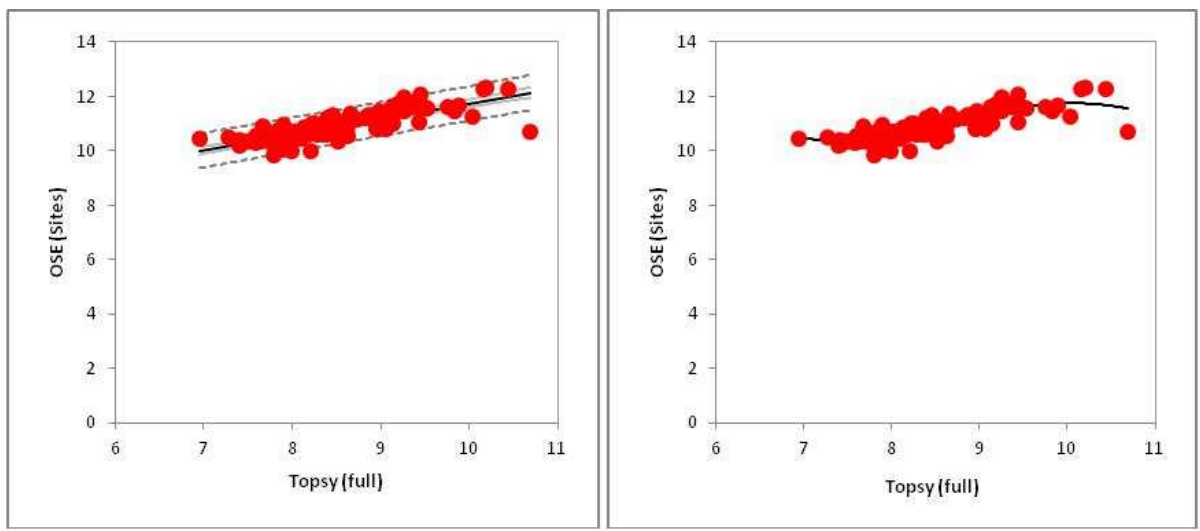

**FIG 5. Linear (left) and nonlinear (right) regression models for USA and Canada.**

**Discussion**

Quantification of tweets

The total number of tweets with an embedded link to each university is high (over 10,000 tweets in 10% of the sample). However, the total amount of tweets collected by university is estimated to be slightly lower than the real number, mainly due to the following considerations:

- Coverage of the API. Topsy announced in September 2013 that its search engine contains all tweets since the birth of Twitter in 2006 (more comprehensive than Twitter's own API). Nonetheless, the data for this research study was collected a few months earlier, in March 2013, when the retrospective coverage may have been slightly lower (Bogers & Björneborn, 2013).

- Public tweets. All data gathered in the Topsy search index comes exclusively from public tweets; private tweets are excluded.

- "Site" search command. The accuracy of this function should be further analyzed. In any case, this aspect statistically affects all universities in equal measure, so its effect is diluted when taking the relative differences between universities.

- Universities with different web domains. Some institutions may have changed their URL since 2006 or have more than one active URL. This is the case of the University of Illinois at Urbana-Champaign (uiuc.edu; illinois.edu), Ohio State University-Main Campus (osu.edu, ohio-state.edu) and the Autonomous University of Barcelona (uab.es, uab.edu, uab.cat).

In this study, only one URL has been used per university, specifically that provided by the Ranking Web of Universities, which, in the event of detecting more than one valid URL, uses the -currently active- URL which performs best for the university, as indicated in their methodology.[5]

Therefore, universities with more than one URL are underrepresented in the study, which suggests that future work should include all variants of localized URLs (both active and former) of the universities in the sample to increase the accuracy of the results. In any case, the number of institutions with this problem is low, and the influence of these alternative URLs in the statistical tests is expected to be not significant.

*Twitter (via Topsy) as an alternative link source to universities*

Correlations, PCA and clusters were calculated in order to study the relationship of links embedded in tweets with other web indicators and to validate Topsy as an alternative source of hyperlink information.

With respect to the correlations observed between the three groups of selected indicators (Topsy, MAJ and OSE), the results are positive and significant. The best correlation observed was that between Topsy (full) and OSE (sites), with a value of r=.80, while it was more moderate for the other indicators. The fact that better results were achieved with OSE (sites) than with OSE (links) reinforces the validity of the first indicator, as already noted by Vaughan and Yang (2012). With regard to the other indicators provided by Topsy (month and week), these are quite stable and may be replaced by the total as well.

On the other hand, the results of the PCA, and the Friedman, Kruskal-Wallis and Nemenyi tests clearly indicate a statistically significant difference between the 7 indicators. These results further reinforce the high correlation value obtained between OSE (sites) and Topsy (full).

*Link prediction*

Regarding the prediction models that were analyzed, the results indicate greater precision in the nonlinear regression (not widely used in link analysis literature), which produced better coefficients of determination. However, application of these models (both linear and nonlinear) gave high error rates in the prediction of links from tweets.

The heterogeneity of university clusters and the behavior of certain universities (outliers) are considered to be the direct causes of these results. Application of the regression models to a specific university environment (in this case USA and Canada) led to a significant reduction in error rates of the models (especially in the linear model) and higher coefficients of determination.

*General limitations*

The number of tweets obtained per URL may be influenced by the following external variables, which should be considered in order to appropriately contextualize the results gathered previously:

- Time of collection. The collection of data was sectional (March 2013), so it may have been influenced by particular, recently concluded events that could have favored certain universities. However, the key indicator is the full tweets indicator, which accounts for such previous events for all universities, so this effect is minimized. In addition, the full tweets indicator achieves high correlation both with weekly and monthly tweets, which corroborates the fact that this factor bore little influence on the correlations obtained. In any case, a future longitudinal analysis should be of interest to check the influence of recent events.

- Self-citation. Universities can receive links from the tweets of their own institutional accounts on Twitter, which is a bias, especially for larger and more technological universities. Topsy's advanced search option offers the possibility of determining the number of self-citations by combining the "site" and "from" commands. For example, the case of Harvard University would proceed as follows:

  <site:harvard.edu from:@harvard>

  However, there is a technical obstacle to excluding self-citations on a massive scale in a large sample of universities due to the difficulty in identifying all existing institutional accounts (not just the general university, but schools, colleges, departments, research groups, libraries, etc.) and combining them into a single accurate query on Topsy.

  In any case, the study of self-citation rate from institutional Twitter accounts to university websites is a topic of interest that should be addressed in future research, although not central to the goals of this work, which is focused on analyzing the possibilities of "Full Tweets" as an alternative or substitute indicator of total number of external links, which in fact include those from Twitter and other social networking sites.

- Hyperlink use motivation within Twitter.

  The reasons for creating a hyperlink within a tweet directed to a university are diverse, although in this study we have considered all equally. Some issues to consider are the following:

  a) The effort to create a hyperlink from a blog or website (you need to be the webmaster and, a few years ago, even have a little knowledge of HTML) is different from the effort required to create it in a tweet (although you need to be a registered user, a simple retweet can generate it automatically).

  b) Hypertextual nature. Links have unique features compared to other indicators obtained through Twitter (such as the number of user mentions or followers) as it is based on the existence of a hyperlink, constituting a transitivity and navigational measurement (Ortega et al., 2014), which differs from other types of mention that may be included in tweets.

c) According to the source of the hyperlink: all tweets are recorded regardless of who generates it (a student, a worker or researcher of any university). Additionally, the area of work or study underlying each of them will influence the creation of hyperlinks, because there are disciplinary differences in scholarly communication on Twitter (Holmberg & Thelwall, 2013), which may introduce an advantage for some institutions.

d) According to the target: all tweets are recorded regardless of whether the target of the link is a PDF file that includes a scientific paper, a slide from a conference, lecture notes, general information about the university, a map, etc.

- Twitter use across countries. The number of active Twitter users per country may introduce a bias for some universities. The statistics portal Statista[20] shows that the United States accounts for 24.3% of Twitter's active users (as of October 2013). Nevertheless, other important countries in the sample (Germany, United Kingdom and China, with 13, 11, and 9 universities respectively) obtain low percentages. The biplot shown in Figure 2 reinforces this bias, since the clusters generated are highly determined by the geographical location-similarity of the universities.

- Outliers. When analyzing the impact of universities on Twitter, it is important to bear in mind the particular behavior of some universities. Universities with very high performance on Twitter were identified (i.e. UNAM, Iowa State University, University of Tokyo and London School of Economics) as were others with much lower performance compared to the total number of links received, notably the University of California-Los Angeles on the one hand, and the Chinese and Taiwanese universities on the other.

In this regard, a distinction must be drawn between cases in which this higher or lower performance is due to a better or worse use of this channel, and cases in which it is due to the observation of specific policies relating to the use of social networking at national level (which explains the results for China and Taiwan).

For universities with very high performance, the possible causes of this effect need to be determined. In this regard, a manual checking of the results provided by Topsy was performed for the 3 main outlier universities of the sample by annotating first the target of the URLs (taking a sample of the first 100 results), then identifying highly used target URLs and finally using these to perform specific queries through the "site" command to check their weight in the results.

This procedure shows how the UNAM owes much of its visibility to the links received by the newspaper La Jornada[21]; Iowa State University, to the IEM weather service (Iowa Environmental Mesonet)[22]; and LSE, to a quality blog.[23]

<site:jornada.unam.mx>; <site:mesonet.agron.iastate.edu>; <site:blogs.lse.ac.uk>

These cases are paradigmatic in the sense they that actually determine the position of these universities in social networks. These platforms undoubtedly generate socially relevant and interesting knowledge or information and are, of course, accessible to an audience that goes beyond the merely academic. This aspect has previously been identified by Orduña-Malea (2013) in the case of the Spanish university system, where platforms like Dialnet (at the University of La Rioja) represent a very high percentage of the total presence of the university on the Net.

**Conclusions**

The use of Twitter as a source of data for webometric analysis would seem to produce promising results. Despite the fact that it represents a small fraction of the total number of links that a university receives, the amount of data received is high and correlates very significantly with link indicators.

However, using it to predict the total number of links still has certain limitations that prevent it being fully recommended for global use. The highly heterogeneous nature of web policies and culture from one country to another, the widespread use of languages, and the specific policies of each university regarding their social media presence lead to universities with very high or very low results; this affects the accuracy of the prediction models. In this regard, the results obtained show how "Full Tweets" -measured via Topsy- could be used as a complementary indicator to external inlinks in the US academic system, but are inadequate if applied on a global basis.

It may therefore be concluded that, although Topsy indicators correlate with other link indicators, measuring the number of tweets with hyperlinks differs somewhat from other link metrics. This means that we are measuring a sub-dimension of hyperlinks marked by the characteristics of the medium (Twitter), which offers another very similar, but complementary, type of information. In this regard, both the progressive adaptation of universities to social media use and presence, and a possible filtering of university outliers could lessen this effect.

**Endnotes**

1. Altavista. Retrieved September 11, 2013, from http://en.wikipedia.org/wiki/AltaVista

2. MajesticSEO. Retrieved September 11, 2013, from http://www.majesticseo.com

3. Ahrefs. Retrieved September 11, 2013, from http://ahrefs.com

4. Open Site Explorer. Retrieved September 11, 2013, from http://www.opensiteexplorer.org

5. Ranking Web of Universities. Retrieved September 11, 2013, from http://www.webometrics.info/en/Methodology

6. Blekko. Retrieved September 11, 2013, from http://blekko.com

7. Blekko SEO Tools. Retrieved September 11, 2013, from https://seo.blekko.com

8. Alexa. Retrieved September 11, 2013, from http://www.alexa.com

9. List of virtual communities with more than 100 million active users. Wikipedia. Retrieved September 11, 2013, from http://en.wikipedia.org/wiki/List_of_virtual_communities_with_more_than_100_million_users

10. Alexa Top Sites. Retrieved September 11, 2013, from http://www.alexa.com/topsites

11. Library of Congress. "Update on the Twitter Archive At the Library of Congress". Retrieved September 11, 2013, from http://www.loc.gov/today/pr/2013/files/twitter_report_2013jan.pdf

12. Track Social. Retrieved September 11, 2013, from http://tracksocial.com

13. Klout. Retrieved September 11, 2013, from http://klout.com/#/pulse/us-universities

14. Social Analytics for Russell Group University Twitter Accounts. Retrieved September 11, 2013, from http://ukwebfocus.wordpress.com/2011/06/28/social-analytics-for-russell-group-university-twitter-accounts

15. Institutional Use of Twitter by Russell Group Universities. Retrieved September 11, 2013, from http://ukwebfocus.wordpress.com/2011/01/14/institutional-use-of-twitter-by-russell-group-universities

16. Topsy. Retrieved September 11, 2013, from http://topsy.com

17. Bits. The New York Times. "If Google Could Search Twitter, It Would Find Topsy". Retrieved September 11, 2013, from http://bits.blogs.nytimes.com/2013/09/04/if-google-could-search-twitter-it-would-find-topsy

18. Crunchbase. Topsy. Retrieved September 11, 2013, from http://www.crunchbase.com/company/topsy-labs

19. Complementary material. Retrieved September 18, 2013, from http://hdl.handle.net/10481/28115

20. Statista. Retrieved September 18, 2013, from http://www.statista.com/chart/1642/regional-breakdown-of-twitter-users

21. La Jornada. Retrieved September 11, 2013, from http://www.jornada.unam.mx

22. Iowa Environmental Mesonet. Retrieved September 11, 2013, from http://mesonet.agron.iastate.edu

23. LSE Blogs. Retrieved September 11, 2013, from http://blogs.lse.ac.uk